\documentclass[useAMS,usenatbib]{mn2e}
\usepackage{graphicx}
\usepackage{lscape}
\usepackage{longtable}
\usepackage{rotating}
\usepackage{colordvi}
\usepackage{color}
\title[Testing gravitational theories...]{Testing gravitational theories using Eccentric Eclipsing  Detached Binaries}
\author[M.~De~Laurentis,\, R.~De~Rosa,\, F.~Garufi and L.~Milano]{M.~De~Laurentis$^{1,2}$, R.~De~Rosa$^{1,2}$, F.~Garufi$^{1,2}$\thanks{E-mail:Fabio.Garufi@na.infn.it}, L.~Milano$^{1,2}$\\
$^{1}$Dipartimento di Scienze Fisiche, Universit\`a di Napoli ``Federico II'',\\
Compl. Univ. di Monte S. Angelo, Edificio G, Via Cinthia, I-80126, Napoli, Italy.
\\
${^2}$INFN Sez. di Napoli, Compl. Univ. di Monte S. Angelo, Edificio G, Via Cinthia, I-80126, Napoli, Italy.}
\begin{document}
\date{Accepted xxxx Yyyyber zz. Received xxxx Yyyymber zz; in original form xxxx Yyyyber zz}
\pagerange{\pageref{firstpage}--\pageref{lastpage}} \pubyear{2012}

\maketitle
\label{firstpage}

\begin{abstract}
In this paper we compare the effects of different theories of gravitation on the apsidal motion of a sample of Eccentric Eclipsing Detached Binary stars. The comparison is performed by using the formalism of the Post-Newtonian parametrization to calculate the theoretical advance at periastron and compare it to the observed one, after having considered the effects of the structure and rotation of the involved stars.\par
A variance analysis on the results of this comparison, shows that no significant difference can be found due to the effect of the different theories under test with respect to the standard General Relativity. It will be possible to observe differences, as we would expect, by checking the observed period variation on a much larger lapse of time. It can also be noticed from our results, that f(R) theory is the nearest to GR with respect to the other tested theories.

\end{abstract}

\section{Introduction}
The problem of the motion of two bodies under their mutual gravitational attraction and the study of binary stellar systems has always been the ideal test bed for the theories of gravitation.
Several authors in the last decades dedicated a lot of work in analyzing, both from the theoretical and the experimental point of wiew, the phenomenon of the periastron precession in binary systems to test various gravitational theories \citep{breen,linsen} as well as to find correction to the Newtonian and General Relativistic behaviour of the systems due to stellar form factors, spin, tides and other phenomena \citep{GimenezClaret,Gimenez,Wolf}.\par
The classical effect of General Relativity (GR) on the  apsidal motion rate at periastron is well known since long time and described by Levi-Civita in a famous paper in 1937 \citep{LeviCivita,Gimenez}. Another possible formulation of the problem, that allows also to test other gravitational theories besides GR, is the use of Parametrized Post Newtonian (PPN) formalism \citep{ThorneWill,Nordtvedt1}.
Using this formalism, the different gravitational theories can be compared side by side on the basis of a set of Post-Newtonian (PN) parameters, the masses, the system major semiaxis and the eccentricity. Thus, using a sample of Eccentric Eclipsing Detached Binary (EEDB) systems, for which masses and orbital parameters are known with sufficient precision, it is possible to compare the apsidal motion rate at periastron $\dot\omega_{Th}$, as expected in the different theories with the observations in order to verify whether the observations can select one theory or another. \par
In the first section we outline the calculation of the   apsidal motion rate at periastron as a function of the Post-Newtonian parameters and give an expression of the PPN for GR, Brans-Dicke (BD) and Nordtvedt (ND) theories \citep{BD,Nordtvedt1}. In the second section, we introduce briefly $f(R)$-theories and calculate the PN parameters for a class of $f(R)$-Lagrangian, see \citep{review} and references therein. 
In the next section we describe the choice of the data sample and present the calculation of $\dot\omega_{Th}$ with its error in the four above mentioned theories.
 Finally, the results obtained using the 'observed' internal second order stellar structure constants (ISC), are compared with those derived by the stellar evolution model, in order to verify whether some relativistic theory can be ruled out. 
 
\section{Advance at periastron}

The calculation of the advance at periastron in a binary system in the frame of the PPN formalism is mainly described in a paper by Breen \citep{breen} that we assumed as reference for this work. In this section, for reader's benefit, we mostly resume the part of Breen's paper that is of interest for the present work.\par
The idea of considering relativistic gravitational tests in terms of a metric expansion is originally based on a work by Shiff \citep{shiff} who expanded the single body metric in terms of the ratio between the geometrized mass $m_g=Gm/c^2$ and the distance $r$:
\[
g_{00}=1-2 \alpha\frac{m_g}{r}+2\beta\left(\frac{m_g}{r}\right)^2
\]
\[
g_{0k}= 0
\]
\begin{equation}
g_{ik}= -\left(1+2\gamma \frac{m_g}{r}\right)\delta_{ik} \;\; i,k = 1,2,3
\end{equation}
Nordtvedt \citep{Nordtvedt1} extended the above metric by writing down a general postnewtonian ($c^{-2}$)
expansion for $n$ moving bodies, introducing four new parameters $\alpha'$, $\alpha''$, $\alpha'''$, $\Delta$ to account for relative velocities and accelerations. Breen \citep{breen} specialized the case for 2 bodies to write the relative acceleration $\mathbf{a_{12}}$ and calculate the angular advance at periastron in terms of the masses and the PN parameters. Writing the acceleration on the plane of the orbit, centered on body 1:
\begin{equation}
\begin{array}{ll}
\ddot r &- r\dot\theta^2 = -\frac{mc^2}{r^2}\left[\alpha-2(\beta+\alpha\gamma)\left(\frac{m_1^2+m_2^2}{mr}\right)+\right.\\
&-2(\alpha+4\alpha\Delta)\left(\frac{m_1 m_2}{mr}\right)+\\
&-\frac{3}{2}\alpha'''\left(\frac{m_1 m_2}{m}\right)\dot r^2+
\gamma\left(\frac{m_1^2+m_2^2}{m}\right)(\dot r^2+r^2\dot\theta^2)+\\
&\left.+ (2\alpha''+4\Delta-\gamma)\left(\frac{m_1 m_2}{m}\right)(\dot r^2+r^2\dot\theta^2)\right]+\\
&+\frac{\dot r^2}{r^2}\left[2(\alpha+\gamma)\left(\frac{m_1^2+m_2^2}{m}\right) +(8\Delta-\alpha'''-\alpha)\left(\frac{m_1 m_2}{m}\right)\right]
\end{array}
 \end{equation}
\begin{eqnarray}
 \nonumber
\frac{1}{r}\frac{d}{dt}(r^2\dot\theta) = \frac{\dot r\dot\theta}{r}\left[2(\alpha+\gamma)\left(\frac{m_1^2+m_2^2}{m}\right)+ \right.\\
\left.+(8\Delta-\alpha'''-\alpha)\left(\frac{m_1 m_2}{m}\right)\right]
\label{eq:accel2}
\end{eqnarray}
where all the masses are geometrized and $m=m_1+m_2$. By integrating eq.(\ref{eq:accel2}), after some manipulation, the advance  at periastron per revolution can be calculated as:
\begin{eqnarray}
\nonumber
\Delta\theta= \frac{2\pi G^2}{c^2 h^2}\left\{(m_1^2+m_2^2)\left[2\alpha(\alpha+\gamma)-\beta\right]+\right.\\
\left.+ m_1 m_2\left[\alpha(8\Delta+2\alpha''-\alpha'''-\gamma-\alpha)-\alpha'\right]\right\}
\label{eq:adv_peri}
\end{eqnarray}
where
\begin{equation}
h^2 = G m a(1-e^2)
\end{equation}
is the square of areolar velocity, $a$ the major semiaxis, $e$ the eccentricity and all the masses are in natural units.\par
The difference for the various theories is in the values of the PN parameters; in this version of the Parametrised Post-Newtonian approximation\footnote{Other versions exist (see e.g. \citep{will}), with a larger number of parameters, where, for GR, only $\beta$ and $\gamma$ are equal to $1$, and the other parameters are $0$. We use the present formulation to be coherent with the notation of \citep{breen}.}, for General Relativity all the parameters are equal to $1$ and the advance at periastron of eq.(\ref{eq:adv_peri}) can be verified to reproduce the "classical" formula by Levi Civita \citep{LeviCivita}:
\begin{equation}
\dot\omega = \frac{6\pi G m}{c^2}\frac{1}{a(1-e^2)}
\label{LeviCivita}
\end{equation}
The above PN parameters, and thus the advance at periastron, can be calculated also for other gravitational theories. For the Brans-Dicke gravitational theory \citep{estabrook}:
\[
\alpha = \alpha{'}= \alpha{'''}= 1
\]
\[
\Delta = \alpha{''}= \left(\frac{3+2\omega}{4+2\omega}\right)
\]
\begin{equation}
\gamma  =  \left(\frac{1+\omega}{2+\omega}\right) 
\end{equation}
where $\omega=5$ is the dimensionless constant of the theory; in the Nordtvedt gravitational scalar-tensor theory \citep{estabrook,Nordtvedt2} $\alpha=\alpha'''=1$, $\alpha''$, $\Delta$ and $\gamma$ are as in Brans-Dicke and:
\begin{eqnarray}
\nonumber
\alpha' = 1+\frac{2\omega'}{(4+2\omega)(3+2\omega)^2}\\
\beta = 1+\frac{\omega'}{(4+2\omega)(3+2\omega)^2}
\end{eqnarray}
being $\omega$ the same as in Brans-Dicke and
\begin{equation}
\omega'=-\frac{1}{2}(3+2\omega)^2
\end{equation}
In the next section we will compute the PPN parameters for $f(R)$ theories.
\section{The PPN parameters for $f(R)$ theories}

From a conceptual point of view, there are no a priori reason to restrict the
gravitational Lagrangian to a linear function of the Ricci scalar R, minimally coupled with matter. Considering higher order terms in R, is the approach of the so called $f(R)$ Extended Theories of Gravitation (ETG) that motivate this approach with considerations coming from cosmology and quantization issues.\par
\par
If one takes into account a more general theory of gravitation, the
calculation of the PPN-limit can be performed following a well
defined pipeline which straightforwardly generalizes the standard
GR case \citep{will}. A significant development in this sense has
been pursued by Damour and Esposito-Farese
\citep{damour1,damour_21,damour_22,damour_23} who approached
the calculation of the PPN-limit of scalar-tensor gravity by
means of a conformal transformation to the standard Einstein frame.
 This scheme  provides several
interesting results up to obtain an intrinsic definition of the parameters
$\gamma$ and $\beta$ in term of the non-minimal coupling function
(see, e.g. \citep{review}). The analogy between scalar-tensor gravity and higher
order theories of gravity has been widely demonstrated
\citep{teyssandier83,schmidtNL,wands:cqg94}.
Scalar-tensor theories and $f(R)$ theories can  be rigorously
compared, after conformal transformations, in the Einstein frame
where both kinetic and potential terms are present.\par
Starting from this analogy, it is possible to extend the definition of the scalar-tensor
PPN-parameters $\gamma$ and $\beta$ \citep{damour1,schimd} to the case of fourth order
gravity, \citep{ppn-noi,capozzMPLA}:
\[
\gamma-1=-\frac{{f''(R)}^2}{f'(R)+2{f''(R)}^2}
\]
\begin{equation}
\label{ppn-R1}
\beta-1=\frac{1}{4}\left(\frac{f'(R)\cdot
f''(R)}{2f'(R)+3{f''(R)}^2}\right)\frac{d\gamma}{dR}.
\end{equation}
If one considers Eqs.(\ref{ppn-R1}) as differential equations, and this hypothesis is reasonable if the
derivatives of f(R) function are smoothly evolving with the Ricci scalar, one can try to
derive a minimal class of f(R) theories, that turns out to be of the form:
\begin{equation}
\label{eq:fourth_order}
f_{\pm}(R) = \frac{1}{12}\frac{1-\gamma}{2\gamma-1}R^3 \pm \frac{1}{2}\sqrt{\frac{1-\gamma}{2\gamma-1}}R^2+R+\Lambda
\end{equation}
where $\Lambda$, at this level, is simply a constant. This expression, though, gives $\beta$ and $\gamma$ that are consistent with GR; alternatively a generic expression such as a small correction to the exponent $1$ of $R$ in GR can be considered, i.e.
\begin{equation}
\label{eq:unemenoeps}
f(R) = R^{1+\epsilon}
\end{equation}
By substituting (\ref{eq:unemenoeps}) into Eqs.(\ref{ppn-R1})  
 one obtains:
\[
\gamma_{\epsilon} = \frac{R^2 + \epsilon^2 (1 + \epsilon) R^{\epsilon}}{R^2 + 2 \epsilon^2 (1 + \epsilon) R^\epsilon}
\]
\begin{equation}
\label{eq:gammabetaeps}
\beta_{\epsilon} =  1 - \frac{(-2 + \epsilon) \epsilon^3 (1 + \epsilon)^2 R^{2 + 2 \epsilon}}{
 4 \left[R^2 + 2 \epsilon^2 (1 + \epsilon) R^{\epsilon}\right]^2 \left[2 R^2 + 3 \epsilon^2 (1 + \epsilon) R^{\epsilon} \right]}
\end{equation}
If one takes the value of $R$ in the solar system $R\sim10^{-24}\;$ $g\;cm^{-3}$ in geometrized units, the result is $\gamma=1/2$ and $\beta = 1$ for any $R\ll\epsilon<1$.

\section{Choosing the systems}
In order to verify whether it is possible to identify the best relativistic theory of gravitation,
the observed apsidal motion of binary stars must be compared with the motion derived in the various theories. The theoretical calculation of the "Newtonian" motion, is complicated by the effects due to the form factor of the stars, their rotation and other intrinsic parameters such as e.g. density and surface temperature. Most of these effects can be summarized by using a limited set of parameters (see \citep{Gimenez}), namely the photometric radii of the components, their mass, the eccentricity of the orbit and the internal structure functions for each component. As we will see in next section the latter functions can be condensed in a sort of average structure function $\bar{k}_2$ that can be both inferred from observational parameters or theoretically calculated from orbital parameters.\par
Thus, among the various binary stars catalogues available in literature, we choose a sample of Eccentric Eclipsing Detached Binary (EEDB) stars such that the period, the eccentricity, the masses of the components, and, possibly, the observed internal structure function are known with a good precision. The EEDB sample we have chosen is shown in Tab.~\ref{table1} and was extracted from the most recent catalogues of eclipsing binaries \citep{Bulut-Demircan,Petrova-Orlov,PetrovaOrlov,DremovaSvechnicov} as well as from the paper by G. Torres \citep{GTorres}.\par
In Tab.~\ref{table1} we report the name of the systems displaying apsidal motion, the classification of the systems with respect to their Roche lobes (Type: D for Detached, SD for Semi-Detached, C for Contact), the observed and theoretical apsidal periods in year $ U_{Obs} $ and  $ U_{Th} $), the orbital sidereal period  $Ps$ in days, the photometric relative radii $r1$ and $r2$ of the binary system components, the orbital eccentricity $e$ and the masses $ M1_{\odot} $ and $  M2_{\odot} $ of the of the binary system components in solar mass unit. For all the data, the error on the least significant digits is reported in parenthesis.
\section{Equations of apsidal motion and Data analysis}
\label{subsec:equations}
To compare the global rates of theoretical and observed apsidal motion we must take into account the individual contributions of each component due to tidal and rotational distortions, and the
general relativistic term $\dot\omega_{\rm Th}$, where the index $Th$ indicates the theory under test (e.g. $\omega_{GR}$ for General Relativity). Assuming that rotation of both components of an eclipsing binary system is perpendicular to the orbital plane, the apsidal motion rate, $\dot\omega$ is given by the following simple relation \citep{Russell-1928,Sterne-1939,Martynov-1971,Kopal-1978}:
\begin{equation}
\label{eq:rate}
\dot\omega= \dot\omega_{cl}+\dot\omega_{Rel} 
\end{equation}
Where $\dot\omega_{cl}$  is the classical Newtonian term and $\dot\omega_{Rel}$ is the relativistic contribution. So the period of periastron rotation in year will be:
\begin{equation}
\label{eq:period}
U(yr)= \frac{360 P(d)}{365\dot\omega}
\end{equation}
where $P(d)$ is the orbital period in days, $\dot\omega$ is in degrees per cycle, 360 is the number of degrees in one cycle and 365 days in one year.
For our purposes the dependance of $\dot\omega_{cl}$ on the Internal second order Structure Constants (henceforth ISC) must be evidenced. It descends from the dependance of the theoretical rate of apsidal motion on the ISC, i.e.: 
\begin{equation}
\label{eq:ISC}
\frac{P}{U_{cl}}=\frac{365\dot\omega_{cl}}{360}=c_{21} k_{21}+ c_{22} k_{22}
\end{equation}
Where the parameters $c_{2i}$ are related to those of the binary system by:
\begin{equation}
\label{eq:Ci}
c_{2i}= \left [\frac{15 m_{3-i}}{m_{i}} g(e)+\left(\frac{\omega_{i}}{\omega_{K}}\right)^{2}\left(1+\frac{m_{3-i}}{m_{i}}\right)f(e)\right]\left({r_{i}}\right)^{5}
\end{equation}
where:
\begin{eqnarray}
\nonumber
f(e)&=&  \left( 1 +  \frac{3}{2} e^{2}  + \frac{1}{8}  e^{4} \right)
\left( 1 - e^{2} \right)^{-5}\\
g(e)&=& \left( 1 - e^{2} \right)^{-2}
\end{eqnarray}  
and the square of the ratio between the actual angular rotational
 velocity $\omega_i$ of the EEDB components to the angular
  keplerian orbital velocity  $\omega_{K}$,
   $\left(\frac{\omega_{i}}{\omega_{K}}\right)^{2}$ was approximated
    according to the relation \citep{Kopal-1978} 
 \[
    \left(\frac{\omega_{i}}{\omega_{P}}\right)^{2} \approx \frac{1+e}{(1-e)^{3}}
 \]  
 being $\omega_{p}$  the angular velocity at periastron and $e$ the orbital eccentricity. The validity of this approximation was tested in the works by \citep{ClaretGimenez} and \citep{GimenezClaret}.
So we have:
\begin{eqnarray}
\label{eq:k2obs}
\bar{k}_{2}&=&\frac{c_{21}  k_{21}+c_{22} k_{22}}{c_{21}+c_{22}}  \\
\bar{k}_{2Obs}&=&\frac{\dot\omega_{cl}}{c_{21}+c_{22}}  
\end{eqnarray}
  It must be noticed that the individual ISC's $k_{2,i}$ cannot be obtained from the observations although they can be interpolated from evolutionary codes like those used in
    \citep{ClaretGimenez1989,ClaretGimenez1992}.\par So we can evaluate a mean model dependent $\bar{k}_{2}$ and a mean observation dependent $\bar{k}_{2Obs}$, and compare them to test the evolution stellar models from the observations of apsidal motion.
For main-sequence stars, $\log{k_2}$ is typically of the order of $-3 \div-2$. 
Now, recalling what observed by Breen \citep{breen}, if the expression in braces at
 eq.(\ref{eq:adv_peri}) is a perfect square, the general expression for the relativistic term $\dot\omega_{Rel}$,
 contributing to the advance at periastron, for the different theories can be written as:
\begin{equation}
\label{eq:Rel}
\dot\omega_{Rel}=K_{Th} G \frac{(\frac{M}{P})^\frac{2}{3}}{c^{2}(1-e^{2})} \\
\end{equation}
where $K_{Th}$ will be:
\begin{equation}
\label{eq:KTh}
K_{Th} = 
\left\{
\begin{array}{ll}
 K_{GR}=3 \rightarrow(General-Relativity)\\\\
 K_{BD}=\frac{19}{7}\rightarrow (Brans-Dicke)\\\\
 K_{ND}=\frac{11}{4} \rightarrow (Nordvedt)\\\\
 K_{f(R)}=\frac{13}{4} \rightarrow(f(R))\\\\
 \end{array}
 \right.
 \end{equation}

So, expressing the total mass of the binary eclipsing system in solar mass units and the period in days, being $c$ the speed of light and $G$ the gravitation constant we obtain:

\begin{equation}
\label{eq:Rel_Num}
\dot\omega_{Rel}=1.8167\cdot10^{-4} K_{Th}  \frac{(\frac{M}{P})^\frac{2}{3}}{(1-e^{2})}
\end{equation}
Now, since we obtain the newtonian term of the apsidal motion rate $\dot\omega_{cl}$ using the observed term  $\dot\omega_{Obs}$ and the relativistic term $\dot\omega_{Rel}$ of the apsidal motion rate, we can write: 
\begin{equation}
\label{eq:key}
\dot\omega_{cl}=\dot\omega_{Obs}-\dot\omega_{Rel}.
\end{equation}
Being $\dot\omega_{Obs}$ fixed, we have that $\dot\omega_{cl}$ will vary according to the the dependance (\ref{eq:Rel_Num}) of $\dot\omega_{Rel}$ on the different theories. So referring to (\ref{eq:KTh}) we will have :
\begin{eqnarray}
\label{eq:comp}
\dot\omega_{f(R)}>\dot\omega_{GR}>\dot\omega_{ND}>\dot\omega_{BD}\\
\nonumber
\dot\omega_{cl_{BD}}>\dot\omega_{cl_{ND}}>\dot\omega_{cl_{GR}}>\dot\omega_{cl_{f(R)}}
\end{eqnarray}
and remembering (\ref{eq:k2obs}) we will have:
\begin{equation}
\label{eq:k2_var}
\bar{k}_{2Obs_{BD}}>\bar{k}_{2Obs_{ND}}>\bar{k}_{2Obs_{GR}}>\bar{k}_{2Obs_{f(R)}}
\end{equation}

\begin{figure}
\center
\includegraphics[width=0.95\columnwidth]{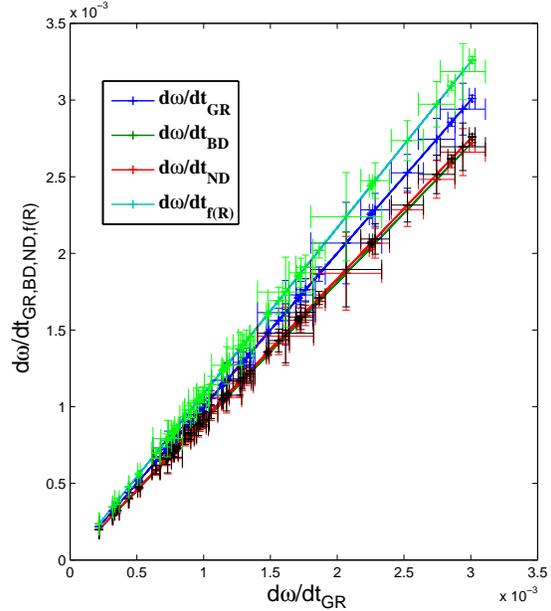}
\caption{$ \dot\omega_{GR}$ vs  $\dot\omega_{GR,BD,ND,f(R)}$ for the different relativistic theories (GR,BD,ND,f(R)): $\dot\omega_{GR}=\dot\omega_{GR}$, $ \dot\omega_{BD}\cong0.92\dot\omega_{GR}$, $\dot\omega_{ND}\cong0.90\dot\omega_{GR}$, $ \dot\omega_{f(R)}\cong1.10\dot\omega_{GR}$}
\label{omegarel}
\end{figure}

\begin{figure}
\center
\includegraphics[width=9 cm, height=9 cm]{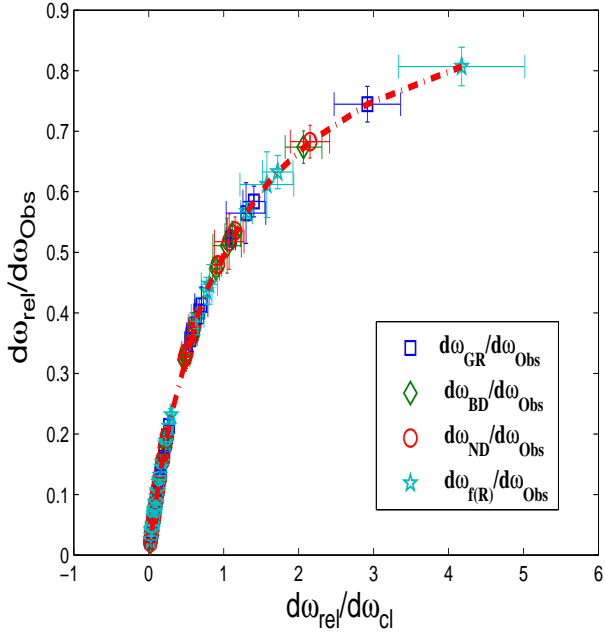}
\caption{Ratio of $\frac{ \dot\omega_{Rel}}{\dot\omega_{cl}}$ vs  $\frac{\dot\omega_{Rel}}{\dot\omega_{Obs}}$ for the different relativistic theories (GR,BD,ND,f(R)). The dashed line represents the relation (see Eq.\ref{eq:key}) $y=\frac{x}{1+x}$, where $y= \frac{\dot\omega_{Rel}}{\dot\omega_{Obs}}$ and $x=\frac{\dot\omega_{Rel}}{\dot\omega_{cl}}$ }
\label{Ratio-figure}
\end{figure}

\begin{figure}
\center
\includegraphics[width=8 cm, height=9 cm]{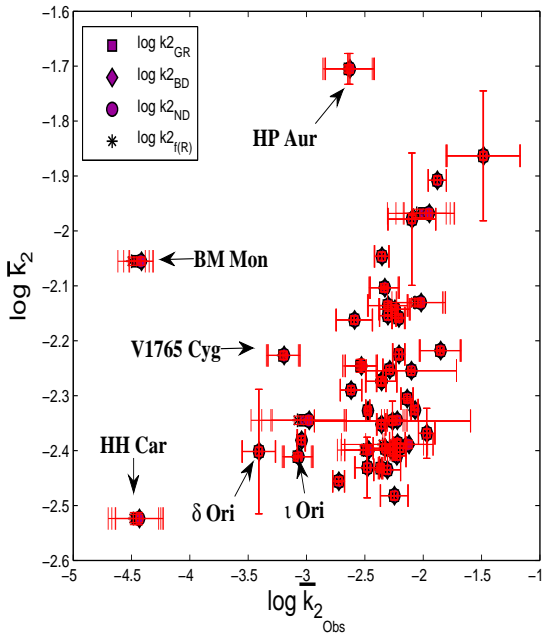}
\caption{Internal second order structure constants (ISC)  $\bar{k}_{2Obs_{Th}}$ vs $\bar{k}_{2}$ for different relativistic terms ($Th\equiv GR,BD,ND,f(R)$)}
\label{ISC-figure}
\end{figure}

\begin{figure}
\center
\includegraphics[width=7 cm, height=7 cm]{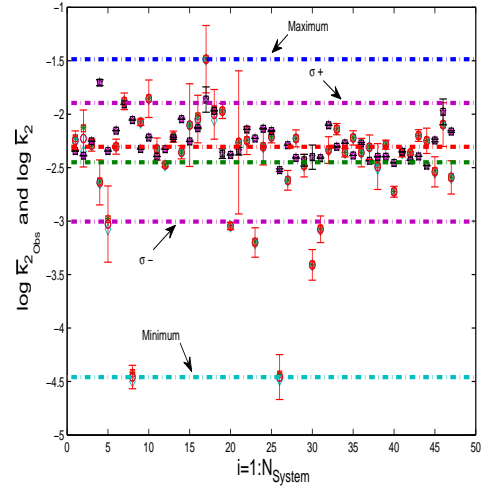}
\caption{Internal second  order structure  constants (ISC)  $\log\bar{k}_{2Obs_{Th}}$ toghether $\log\bar{k}_{2}$ for different relativistic terms ($Th\equiv{GR,BD,ND,f(R)}$). The mean value (green dotted line), the median (red dotted line) and the standard deviation lines (magenta dotted lines) toghether with the maximum (blue dotted line) and minimum (cyan dotted line) of the whole sample of $\log\bar{k}_{2}$ and $\log\bar{k}_{2_{Obs}}$ are also shown (min:$ -4.46 $ ,max: $-1.49$,mean$\pm{\sigma}$: $2.45\pm{0.56}$, median: $-2.30$)}
\label{ISCGlob-figure}
\end{figure}

 In Fig.~\ref{omegarel} we show the trend of the apsidal motion rate $\dot\omega_{GR}$ vs $\dot\omega_{GR,BD,ND,f(R)}$. Recalling (\ref{eq:Rel_Num}) we can write: $\dot\omega_{BD}\cong0.92\,\dot\omega_{GR}$, $\dot\omega_{ND}\cong0.90\,\dot\omega_{GR}$, $ \dot\omega_{f(R)}\cong1.10\,\dot\omega_{GR}$.
It is interesting to note that, according to the values of $K_{Th}$, $f(R)$ theory gives a relativistic contribution that is slightly higher than GR, whilst BD and ND are slightly smaller. The GR relativistic contribution to apsidal motion appears to be approximately the average among the different theories. 
It is also evident that there is no significant difference among the theories under test within the errors; nonetheless, significant differences could be found for massive systems with high orbital eccentricities and short orbital period, being $K_{Th}$ in Eq.(\ref{eq:Rel_Num}) a sort of amplification factor of each relativistic theory.
 In Fig.~\ref{Ratio-figure} the ratio of $\frac{ \dot\omega_{Rel}}{\dot\omega_{cl}}$ vs  $\frac{\dot\omega_{Rel}}{\dot\omega_{Obs}}$ is shown for the different relativistic theories (GR, BD, ND, f(R)). Defining $y= \frac{\dot\omega_{Rel}}{\dot\omega_{Obs}}$ and $x=\frac{\dot\omega_{Rel}}{\dot\omega_{cl}}$, from Eq.(\ref{eq:key}) the relation $y=\frac{x}{1+x}$ holds. The dashed line in Fig.~\ref{Ratio-figure} represents this last relation, and it is evident that, within the error bars, all the points lie on the line. This shows that no significant differences among the theories can be noticed within the errors. 
In Fig.~\ref{ISC-figure} we show $\log\bar{k}_{2_{Obs}}$ vs $\log\bar{k}_{2}$; it worths noticing that $\bar{k}_{2}$ according to Eq.(\ref{eq:k2obs}) depends on the stellar evolution model but not on the relativistic theory. It is evident that some systems deviate from the nearly common trend of  $\log\bar{k}_{2_{Obs}}$ vs $\log\bar{k}_{2}$.
 In Fig.~\ref{ISCGlob-figure} all the theoretical and observed (according to the different theories) values of $\log\bar{k}_{2}$ and $\log\bar{k}_{2_{Obs}}$ are shown. The mean value (green dotted line), the median (red dotted line) and the standard deviation lines (magenta dotted lines) toghether with the maximum (blue dotted line) and minimum(cyan dotted line) of the whole sample of $\log\bar{k}_{2}$ and $\log\bar{k}_{2_{Obs}}$ are also shown on the plot. Also in this graph there are systems that deviate for more than one standard deviation with respect to the mean value.  We postpone to the next section the discussion of these results. To perform a quantitative analysis, we show the results obtained from our computations in Tables~\ref{omegap}, \ref{omegap-ratio} and \ref{k2table}: in Table~\ref{omegap} we show the apsidal motion rate $\dot\omega_{Rel}$ and $\dot\omega_{Obs}$, in Table~\ref{omegap-ratio} we show the ratio of $\frac{\dot\omega_{Rel}}{\dot\omega_{Obs}}$ and $\frac{\dot\omega_{Rel}}{\dot\omega_{cl}}$  and finally in Table~\ref{k2table} the ISC, $\log k_{2i}$  and $\log\bar{k}_{2_{Obs}}$ toghether with $\log\bar{k}_{2}$ for different relativistic terms (GR,BD,ND,f(R)). 


\section{Discussion and Conclusion}
\begin{figure*}
\begin{tabular}{|c|c|}
\includegraphics[width=8 cm, height=6 cm]{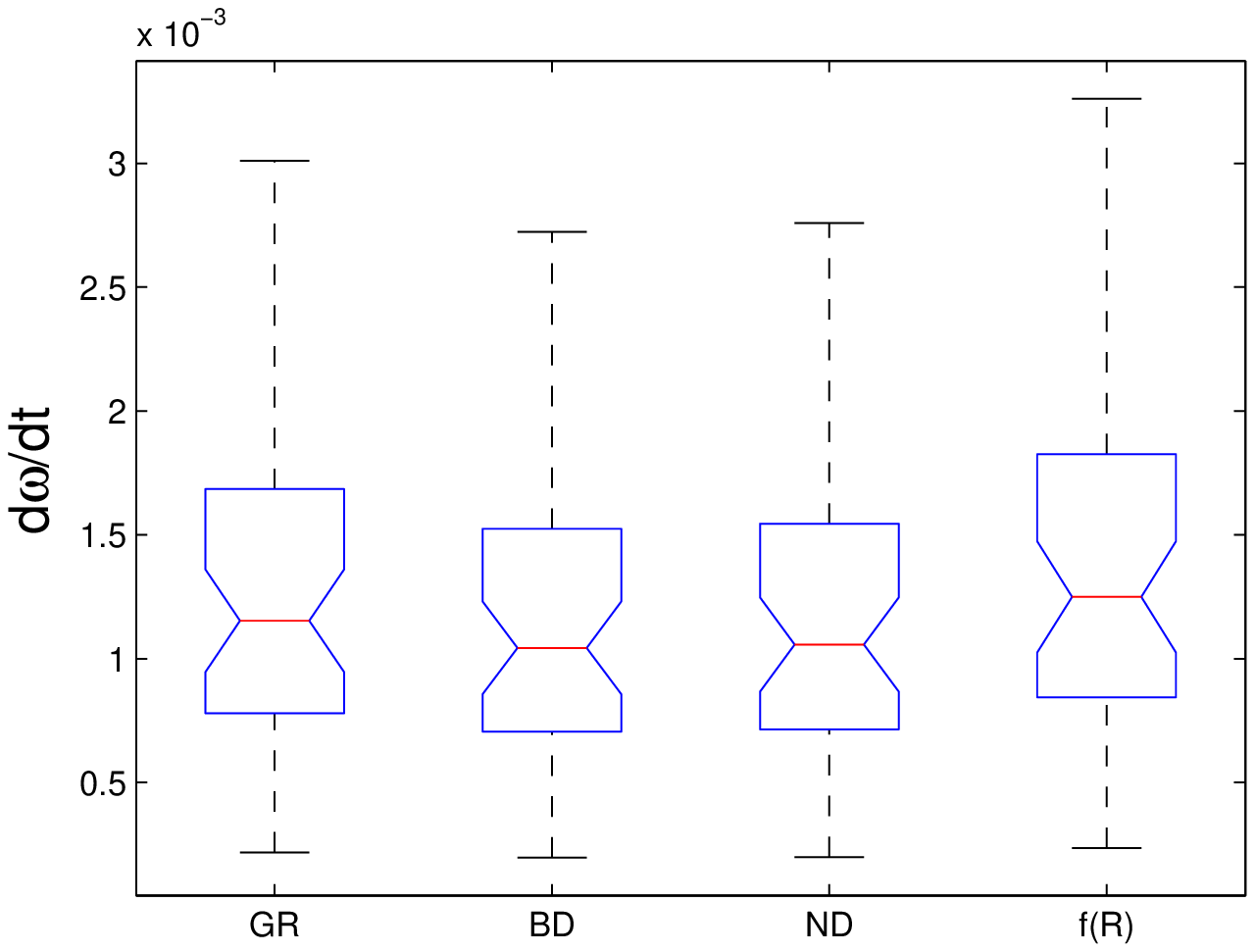}
\includegraphics[width=8 cm, height=6 cm]{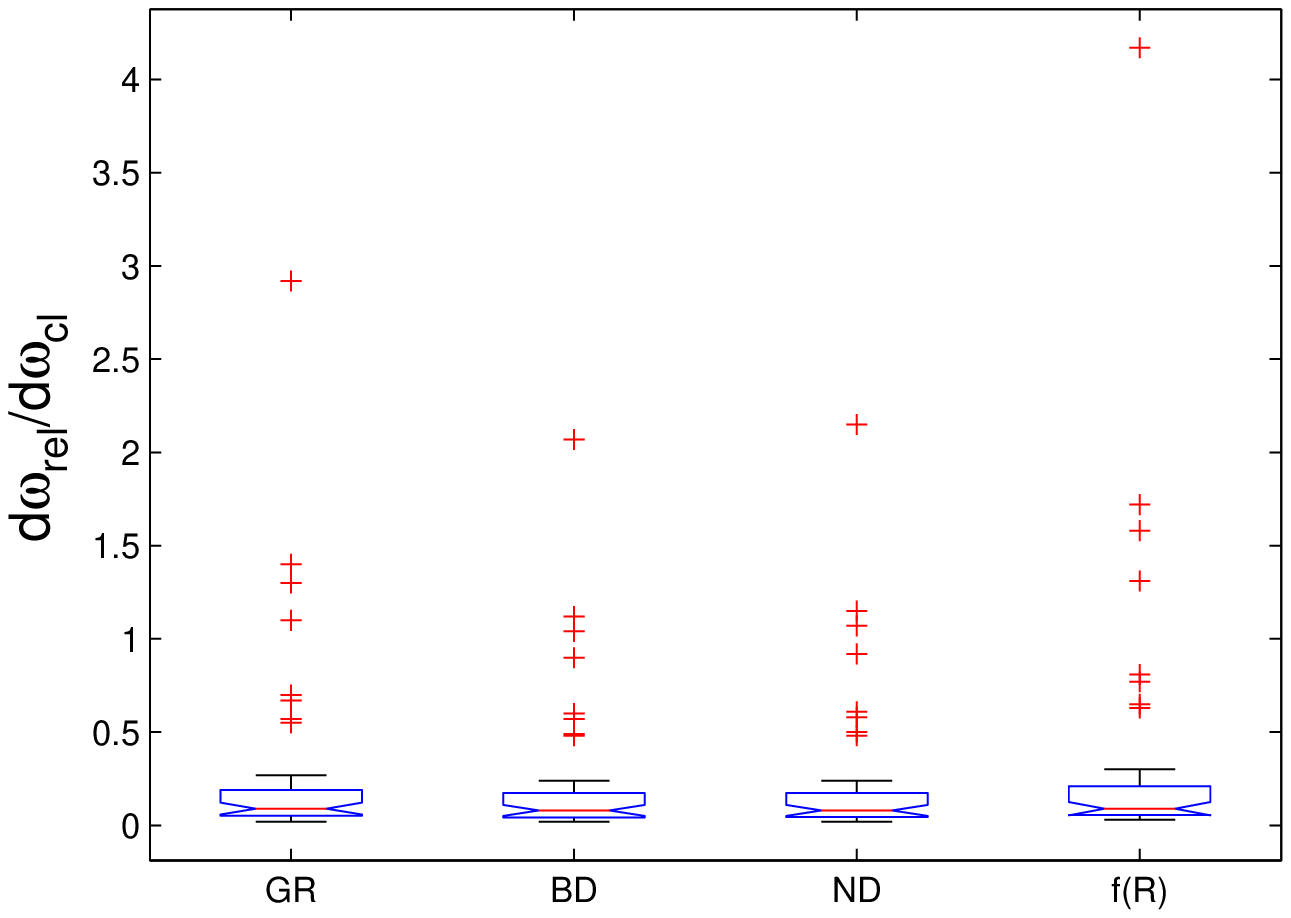}
\tabularnewline 
\includegraphics[width=8 cm, height=6 cm]{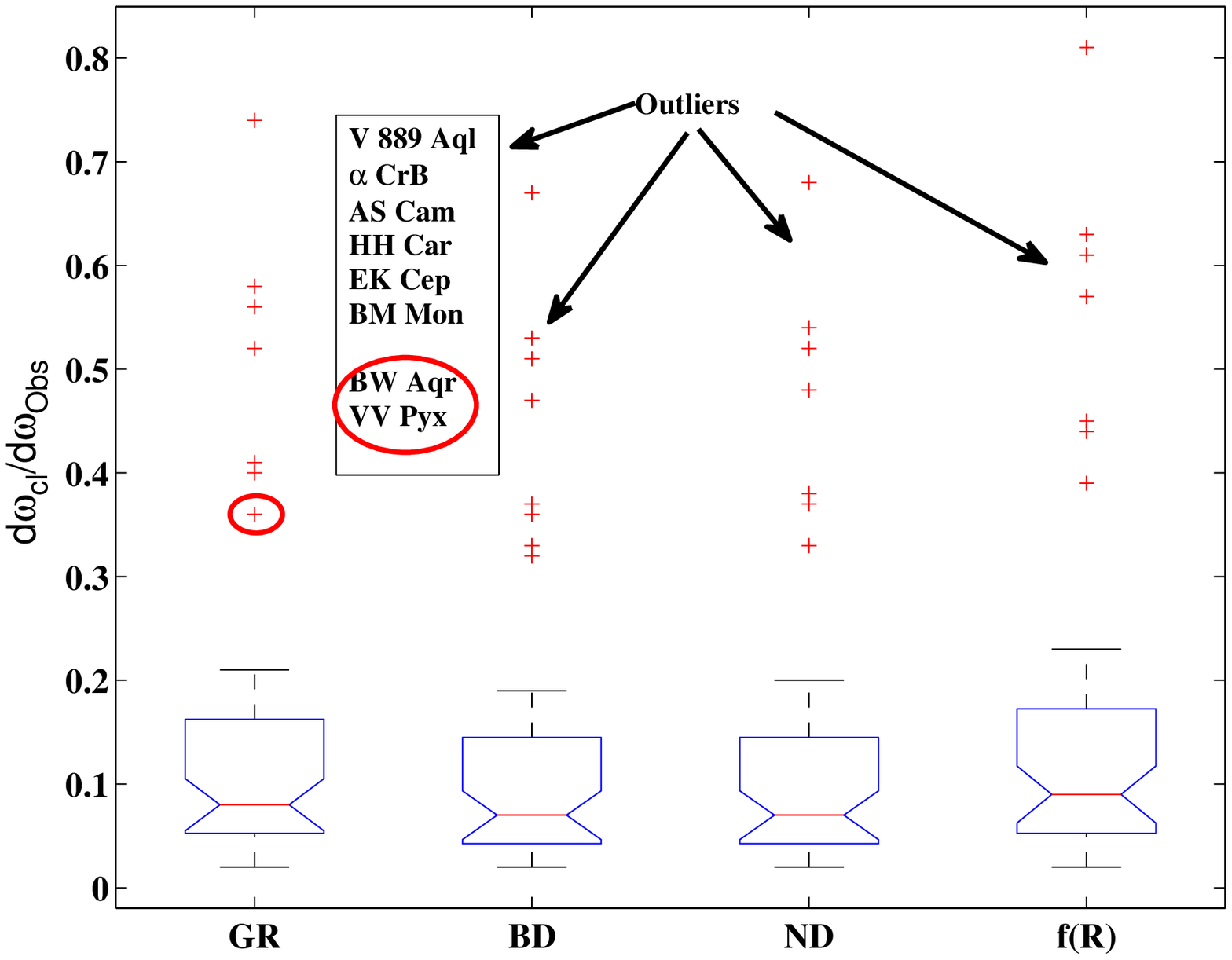} 
\includegraphics[width=8 cm, height=6 cm]{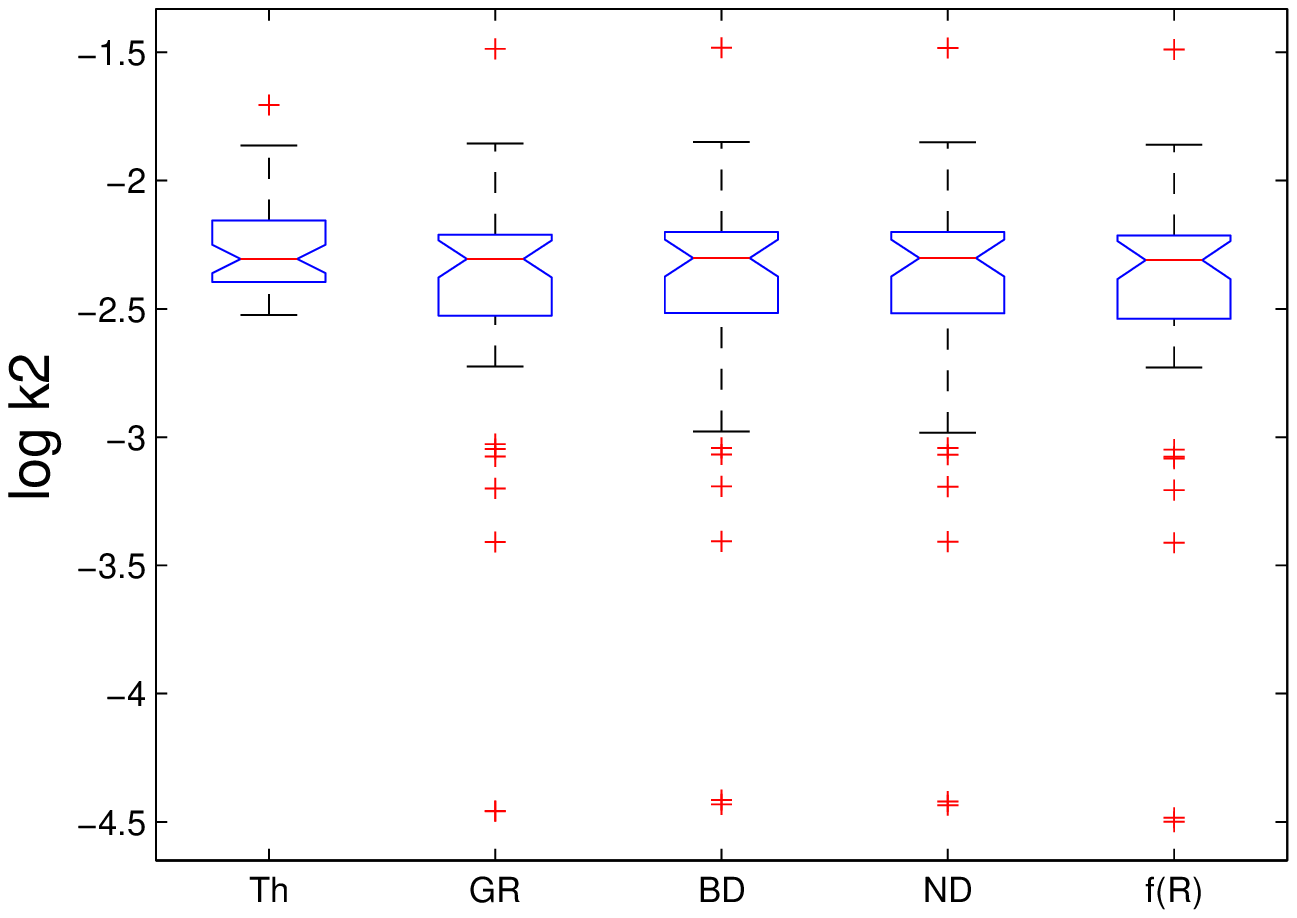}
 \tabularnewline
\end{tabular}
\caption{Results of the one-way variance analysis performed on the groups relative to the different theories under test: $\dot\omega_{GR,BD,ND,f(R)}$ (top left), $\frac{\dot\omega_{Rel}}{\dot\omega_{cl}}$ (top right), $\frac{\dot\omega_{cl}}{\dot\omega_{Obs}}$(bottom left) and the groups constituted by $\log(\bar{k}_{2})$ toghether with $\log(\bar{k}_{2Obs})$ (bottom right). The boxes represent the statistic in the following way: the middle line is the median; the bottom and top sides are the first and third quartiles, the thin lines departing from the box reach the minimum and maximum values not considered as outliers; the crosses are the outliers.}\label{Stat_Ratio-Obs-figure}
\end{figure*}

The key idea we followed to try to constrain different relativistic theories using data coming from apsidal motion rate of EEDB was based on the fact that varying the relativistic term of the apsidal motion rate according to the different relativistic theories, the classical Newtonian term will vary accordingly. In this way we can also test if there is an improved agreement among the observed second order mean stellar structure constants and those obtained from the different relativistic aspidal motion rate terms. These tests are based on the fact that if no significant difference among the theories is found within the errors from the apsidal motion data, we can conclude that, if a difference exists, it is masked by the present uncertainty on the knowledge of the ISC as they are determined both by the models and the observations
 In  Fig.~\ref{Stat_Ratio-Obs-figure}, we show the results of the one-way variance analysis performed on the groups relative to the different theories under test; using the technique of the boxplot \citep{Nelson-1989}, we describe, concisely, the distributions of  $\dot\omega_{rel}$ (top left), $\frac{\dot\omega_{Rel}}{\dot\omega_{cl}}$ (top right), $\frac{\dot\omega_{cl}}{\dot\omega_{Obs}}$ (bottom left) and the groups constituted by $\log(\bar{k}_{2})$ and $\log(\bar{k}_{2_{Obs}})$ (bottom right) for the different theory samples. We must note that the medians of the distributions are not different within the errors, whereas, by the positions of their percentiles, we observe an asymmetry and the presence of eight outliers i.e.: V889 Aql, $\alpha$ Crb, AS Cam, HH Car, EK Cep, BM Mon, BW Aqr, VV Pyx, well known from the literature. The presence of $\approx{17\%}$ outliers means that the mean values of the sample distributions are biased.
 \begin{table}
\center
\caption{
Results of variance analysis for $\dot\omega_{Rel}$,$\frac{\dot\omega_{Rel}}{\dot\omega_{cl}}$, $\frac{\dot\omega_{cl}}{\dot\omega_{Obs}}$ and $\log\bar{k}_{2_{Obs}}$  computed for different values of relativistic terms: GR, BD, ND, f(R). The samples of $\log\bar{k}_{2_{Obs}}$ were tested toghether with the sample of $\log\bar{k}_{2}$. It is evident from the values of the probability $p > F$ much greater than  $5\%$  that the null hypothesis, for the mean we tested, is significant. }
\begin{tabular}{|l|l|l|}
\hline
\hline
& $F_{Fisher} $ &	$Prob>F$ \\ \hline \hline
$\dot\omega_{Rel}$ &  1.13 &  0.34 \\ \hline
$\frac{\dot\omega_{Rel}}{\dot\omega_{cl}}$ & 0.45  & 0.72 \\ \hline
$\frac{\dot\omega_{cl}}{\dot\omega_{Obs}}$ & 0.28  & 0.84 \\ \hline 
$\log\bar{k}_{2_{Obs}}$ & 0.72	&  0.58 \\ \hline \hline
\\
\end{tabular}
\label{Stat_Table}
\end{table}
 
 We  performed a variance analysis for the mean on the groups of observed and theoretical values, to ascertain, in a quantitative way, if there are significant differences among the groups obtained for the tested relativistic theories. The results for the $F$ statistic \citep{Hogg-1987} are shown in Tab.~\ref{Stat_Table} and, due to the values of the probability $p > F$ much greater than  $5\%$, the null hypothesis that there is no significant difference among the relativistic theories in spite of the outliers is confirmed. Of course, the present results could be heavily dependent on the large uncertainties on the $c_{2i}$, through which ISC are determined. There are also other factors, like the hypothesis of syncronization between orbital and rotational period of the binary components, unrevealed presence of a third body, or tidal effects not properly taken into account. In our opinion, an interesting result is that the relativistic contribution to the apsidal motion coming from GR theory is approximately a mean among the different theories through the $K_{Th}$. Now, considering $K_{Th}$ as a sort of amplification factor of each relativistic theory,
 we will have that significant differences could be found for systems with large orbital eccentricities and high total mass $M_{t}$ to orbital  period $P$ ratio $\frac{M_{t}}{P}$. 
\\\\\\\\\\\\\\\

\section*{Acknowledgments}
We warmly thank the referee for the many suggestions that made this work more readable and formally correct.

\begin{table*}
\center
\caption{
Data of the Eccentric Eclipsing Detached Binary (EEDB) sample, displaying apsidal motion, used in this paper. 
We report the name of the systems displaying apsidal motion, the classification of the systems with respect to their Roche lobes (Type: D for Detached, SD for Semi-Detached, C for Contact), then the observed and theoretical apsidal periods in year $ U_{Obs} $ and  $ U_{Th} $), the orbital period  $Ps$ in days, the photometric relative radii r1 and r2 of the binary system components, the orbital eccentricity $e$ and the masses $ M1_{\odot} $ and $  M2_{\odot} $ of the of the binary system components in solar mass unit are shown. For all the data, the error affecting the least significant digits is reported in parenthesis}
\label{table1}
\begin{tabular}{|l|l|l|l|l|l|l|l|l|l|}
\hline
\hline
  Name	&   Type   &    $ U_{Obs}$(yr)   &   $ U_{Th}$(yr)  &  Ps(d)  &  r1   &   r2	  &   $e$   &  $ M1_{\odot} $ & $  M2_{\odot} $  \\
  \hline
  \hline
	BW Aqr	 & 	D	 & 7400(900)	 & 	 8662(217)	 & 	 6.719695(3) &	0.097(2)	  & 	 0.084(2)  & 	 0.17(1)	& 1.49(2)	  & 	1.39(2)	 \\ \hline
V889 Aql	  &  	D	  & 	23200(3500)	&  	 2557(182)	  &  	11.1207937(25)	& 0.0582(5)	 &  	0.0524(5)	  &  	0.375(4)	& 2.4(2)	  &  	2.2(2)	 \\ \hline
 	V539 Ara &  	D	  &  	150(15)	  &  	141(8)	  &  	3.1690854(12) &	 0.220(4)		 &  	0.167(4)	  &  	0.053(10) &	6.24(7)	&  	5.31(6)
 	 \\ \hline
 	GL Car	  &  	D	  &  	25(3)	  &  	 27(7)	  &  	2.4222308(8)	&  0.2204(60)	  &  	 0.2094(60)	  &  	0.1457(10)	& 13.5(1.4)	  &  	 13.0(1.4)	 \\ \hline
 HH Car	  &  	SD	  &  	660(66)	  &  	 255(11)	  &  	3.231553(3) &	 0.21(1)	  &  	 0.368(3)	  &  	0.16(2) &	17(1.7)	 &  	 14(1.4)	 \\ \hline
 	QX Car	  &  	D	  &  	361(6)	  &  	 1097(98)	  &  	4.4779754(2)	& 0.144(3)	  &  	0.136(3)	  &  	0.278(3)  &	9.27(12)	 &  	8.48(12)	 \\ \hline
 	AR Cas	  &  	D	  &  	922(92)	  &  	 74(26)	  &  	6.066317(49)	& 0.1633(20)	  &  	 0.0639(64)	  &  	0.210(20)  &	6.7(7)	  &  	 1.9(2)	 \\ \hline
 	OX Cas	  &  	D	  &  	40(2)	  &  	 114(23)	  &  	2.489345(36) &		 0.2550(120) &  0.247(18)	  &	0.042(2)	 &	11(1.1) &  	 10.30(1.03)\\ \hline
 	PV Cas	  &  	D	  &  	94(2)	  &  	 144(46)	  &  	 1.7504697(14)	 &	 0.2083(13)	  &  0.2121(18)	  &  	0.0320(10)	 &	2.76(6)	 &  	 2.81(5)	 \\ \hline
 	KT Cen	  &  	D	  &  	260(20)	  &  	 198(86)	  &  	 4.1304380(1)    &		 0.171(1)	  &  	 0.159(1)	  &  	0.225(5)  &		5.3(5)	 &  	 5.0(5)	 \\ \hline
 	V346 Cen	  &  	D	  &  	321(16)	  &  	94(2)	  &  	 6.3219156(20)	 &	 0.211(4)	 &  0.107(2)	  &  	0.288(3)	 &	11.8(1.4)	 &  	8.4(8)	 \\ \hline
 	CW Cep	  &  	D	  &  	46(39)	  &  	 2920(698)	  &  	 2.7291396(18)	 &	 0.235(5)	  &  	0.214(1) &  	0.0293(6)  &	11.82(0.14) &  	11.09(14)	 \\ \hline
 	EK Cep	  &  	D	  &  	4100(1200)	 &  	1855(76)	  &  	4.427796(3)  &		 0.095(3)	  &  	0.079(3)	  &  	0.109(3) &	 2.03(2)	  &  1.12(1)	 \\ \hline
 	NY Cep	  &  	D	  &  	1300(800)	  &  	56728(397)	  &  	 15.27566(1)	 &	0.093(10)	   &  	0.074(7)	  &  	0.48(2)  &		13(1)   &  	9(1)	 \\ \hline
 	$\alpha$ Crb	  &  	D	  &  	46000(8000)	  &  	277(14)	  &  	 17.3599002(13)	&		0.071(7)	  &  	0.021(1)	  &  	0.371(5)  &		 2.58(4)	  &  	 0.92(2)	 \\ \hline
 	Y Cyg	  &  	D	  &  	48(2)	  &  	 186(19)	  &  	 2.996846(20) &		 0.211(10)	  &  	 0.199(9)	  &  	0.1458(2)	 &	17.5(4)	 &  	 17.3(3)	 \\ \hline
 	V380 Cyg	  &  	D	  &  	1395(32)	 &  	271(22)	  &  	 12.425719(14)  &		0.271(4)	 &  	0.068(2)	  &  	0.2183(51)	 &	12.1(3)	 &  	7.3(3)	 \\ \hline
 	V453 Cyg	  &  	D	  &  	71(3)  &  	664(72)	  &  	 3.889825(18)  &		 0.28(10)	 &  	0.174(5)	  &  	0.019(5)	 &	13.9(7)	 &  	10.7(6)	 \\ \hline
 	V477 Cyg	  &  	D	  &  	350(10)	  &  	17(3)	  &  	 2.346978(1)	 &	 0.1441(20)	 &  	0.1167(13)	  &  	0.307(3)	 &	1.79(12)	 &  	1.35(70)	 \\ \hline
 	V1765 Cyg	  &  	D	  &  	1930(150)	  &  	810(405)	  &  	 13.373415(10) &		0.257(20)	  &  	0.084(8)	  &  	0.315(15)	 	&	 23.5(1)	  &  	 11.7(5)	 \\ \hline
 	57 Cyg	  &  	D	  &  	203(4)	  &  	 2490(105)	  &  	 2.85480(1)  &		 0.190(20)	  &  	0.160(20)	  &  	0.139(14)	 &	5.54(55)	 &  	4.92(49)	 \\ \hline
 	RU Mon	  &  	D	  &  	348(15)	  &  	 4(1)	  &  	 3.5846513(8)	 &	 0.136(4)	  &  	 0.122(4)	  &  	0.385(5)    &		3.60(40)	  &  	 3.33(0.33)	 \\ \hline
 	BM Mon	  &  	C	  &  	168(34)	  &  	 50(35)	  &  	 1.244951(4)  &		 0.50(5)	  &  	 0.264(30)	  &  	0.18(5)  &		11.7(1.2)	  &  	 3.15(32)	 \\ \hline
 	GM Nor	  &  	D	  &  	90(15)	  &  	 27(2)	  &  	 1.884577(10)	 &	 0.265(3)	  &  	 0.177(5)	  &  	0.045(2)  &		2.2(3)	  &  	 1.8(2)	 \\ \hline
 	U Oph	  &  	D	  &  	21(3)	  &  	 121(10)	  &  	 1.6773458(4)	 &	 0.269(3)	  &  	 0.237(7)	  &  	0.0031(3)   &		5.16(10)	 &  	 4.6(6)	 \\ \hline
 	V451 Oph	  &  	D	  &  	180(30)	  &  	10(2)	  &  	 2.19659700(12)	 &	 0.2155(20)	 &  	0.1655(20)	  &  	0.0125(15)	 &	2.78(60)	 &  	2.36(5)	 \\ \hline
 	$\delta$ Ori	  &  	D	  &  	227(37)	  &  	201(145)	  &  	 5.7325(1)    &		0.43(2)	 &  	0.25(4)	  &  	0.089(1)  &		23(2)	 &  	9(1)	 \\ \hline
 	$\iota$ Ori	  &  	D	  &  	2400(180)	  &  	5184(1483)  &  	29.13376(1) &		 0.103(10)	  &  	 0.062(6)	  &  	0.764(9)  & 38.9(9.7)	  &  	 18.9(4.7)	 \\ \hline
 \hline
 \\
 \\
\end{tabular}

\end{table*}

 	\begin{table*}
\center
\addtocounter{table}{-1}
\caption{continued}
\begin{tabular}{|l|l|l|l|l|l|l|l|l|l|}
\hline
\hline
  Name	&   Type   &    $ U_{Obs}$(yr)   &   $ U_{Th}$(yr) &  Ps(d)  &  r1   &   r2	  &   $e$   &  $ M1_{\odot} $ & $  M2_{\odot} $  \\
  \hline
  \hline
 	FT Ori	  &  	D	  &  	481(19)  &  	 110(26)	  &  	 3.1503919(3)	 &	 0.124(13)	  &  	 0.118(12)	  &  	0.4046(15) &		2.5(3)	 &  	 2.3(3)	 \\ \hline
 	AG Per	  &  	D	  &  	76(6)	  &  	 122(7)	  &  	 2.0287298(100)	 &	 0.2045(45)	  &  	 0.1779(45)	  &  	0.071(10)  &		5.36(16)	 &  	 4.90(13)	 \\ \hline
 	IQ Per	  &  	D	  &  	119(9)	  &  	 52(1)	  &  	 1.74356210(8)	 &	 0.231(2)	  &  	 0.142(3)	  &  	0.076(4)	 &	3.51(4)	  &  	 1.73(2)	 \\ \hline
 	${\zeta}$ Phe	  &  	D	  &  	44(7)	  &  	142(36)	  &  	 1.669770(26) &		 0.2583(14)	 &  	0.1678(21)	  &  	0.0113(20)  &		3.93(4) &  	 2.55(3) \\ \hline
 	KX Pup	  &  	D	  &  	170(30)	  &  	 24(1)	  &  	 2.146795(2)	 &	 0.205(3)	  &  	 0.14(3)	  &  	0.153(12)	 &	2.5(3)	  &  	 1.8(2)	 \\ \hline
 	NO Pup	  &  	D	  &  	37(2)	  &  	 811(16)	  &  	 1.2569966(10)   &		 0.253(10)	  &  	 0.177(10)	  &  	0.1255(10)	 &	2.88(10)	 &  	 1.50(50)	 \\ \hline
 	VV Pyx	  &  	D	  &  	3200(1000)	 &  	665(287)  &  	 4.5961801(50)  &		 0.1156(10)	 &  	0.1150(10)	  &  	0.0956(9)   &		2.09(8)	 &  	2.09(8)	 \\ \hline
 	YY Sgr	  &  	D	  &  	297(4)	  &  	 114(41)	  &  	 2.6284738(6)	 &	 0.1643(12)	  &  	 0.149(3)	  &  	0.1575(7)  &		3.23(32)	 &  	 3.03(30)	 \\ \hline
 	V523 Sgr	  &  	D	  &  	203(1)	  &  	229(142)	  &  	 2.3238131(4) &		 0.229(2)	 &  	0.157(6)	  &  	0.162(10)	 &	1.45(15)	 &  	 1.42(14)	 \\ \hline
 	V526 Sgr	  &  	D	  &  	156(3)	  &  	342(8)	  &  	 1.9194118(8) &		 0.1841(4)	 &  	0.152(1)	  &  	0.2194(4)	 &	2.40(24)	 &  	1.85(19)	 \\ \hline
 	V1647 Sgr	  &  	D	  &  	593(7)	  &  	611(186)	  &  	 3.2827950(20)	 &	0.1226(10)	 &  	0.1116(10)	  &  	0.413(5)	 &	2.19(4)	 &  	1.97(3)	 \\ \hline
 	V760 Sco	  &  	D	  &  	40(3)	  &  	5(1)	  &  	 1.7309338(12)	 &	 0.234(5)	 &  	0.205(8)	  &  	0.0265(10)   &		4.98(9)	 &  	4.62(7)	 \\ \hline
 	AO Vel	  &  	D	  &  	57(2)	  &  	 50(175)	  &  	 1.5846212(7)   &		 0.214(6)	  &  	 0.193(6)	  &  	0.0761(17)	 &	4.4(1.2)	  &  	 3.6(1.0)	 \\ \hline
 	EO Vel	  &  	D	  &  	1600(400)	  &  	1459(136)	  &  	 5.329675(5)	 &	0.141(1)	   &  	0.135(1)	  &  	0.208(4)	 &	3.2(3)	   &  	3.2(3)	 \\ \hline
 	HR 8384	  &  	D	  &  	94(15)	  &  	 158(100)	  &  	 2.99000(1)  &		 0.185(20)	  &  	 0.168(20)	  &  	0.26(15)  &		4.56(50)	 &  	 3.93(40)	 \\ \hline
 	HR 8800	  &  	D	  &  	143(17)	  &  	 1101(87)	  &  	 3.3380(10)	 &	 0.260(30)	  &  	0.16(2)	  &  	0.2410(211)  &		10.3(1.0)	 &  	4.50(45)	 \\	
 \hline
 \hline
 \\
\end{tabular}
\end{table*}

\begin{table*}
\center
\caption{
Apsidal motion rate $\dot\omega_{Rel}$ for different relativistic terms (Rel = GR,BD,ND,f(R)) and $\dot\omega_{Obs}$}

\begin{tabular}{|l|l|l|l|l|l|l|}
\hline
\hline
Name	&	Type	&	$\dot\omega_{GR}$	&	$\dot\omega_{BD}$	&	$\dot\omega_{ND}$	&	$\dot\omega_{f(R)}$	&	$\dot\omega_{Obs}$	\\ \hline \hline
BW Aqr	&	D	&	0.000319	&	0.000289	&	0.000292	&	0.000346	&	0.000896		\\ \hline
V889 Aql	&	D	&	0.000352	&	0.000319	&	0.000323	&	0.000381	&	0.000473		\\ \hline
V539 Ara	&	D	&	0.001294	&	0.001171	&	0.001187	&	0.001402	&	0.020838		\\ \hline
HP Aur	&	D	&	0.000774	&	0.000700	&	0.000709	&	0.000838	&	0.003626		\\ \hline
AS Cam	&	D	&	0.000796	&	0.000720	&	0.000730	&	0.000863	&	0.001410		\\ \hline
EM Car	&	C	&	0.003010	&	0.002724	&	0.002759	&	0.003261	&	0.080179		\\ \hline
GL Car	&	D	&	0.002744	&	0.002483	&	0.002515	&	0.002973	&	0.094728		\\ \hline
HH Car	&	SD	&	0.002525	&	0.002285	&	0.002315	&	0.002736	&	0.004829		\\ \hline
QX Car	&	D	&	0.001479	&	0.001338	&	0.001356	&	0.001603	&	0.012234		\\ \hline
AR Cas	&	D	&	0.000720	&	0.000651	&	0.000660	&	0.000779	&	0.006489		\\ \hline
OX Cas	&	D	&	0.002284	&	0.002066	&	0.002094	&	0.002474	&	0.061690		\\ \hline
PV Cas	&	D	&	0.001180	&	0.001068	&	0.001082	&	0.001279	&	0.018367		\\ \hline
KT Cen	&	D	&	0.001056	&	0.000955	&	0.000968	&	0.001144	&	0.015669		\\ \hline
V346 Cen	&	D	&	0.001289	&	0.001166	&	0.001182	&	0.001397	&	0.019425		\\ \hline
CW Cep	&	D	&	0.002253	&	0.002039	&	0.002065	&	0.002441	&	0.059056		\\ \hline
EK Cep	&	D	&	0.000440	&	0.000398	&	0.000403	&	0.000476	&	0.001065		\\ \hline
NY Cep	&	D	&	0.000903	&	0.000817	&	0.000828	&	0.000978	&	0.011590		\\ \hline
$\alpha$ Crb	&	D	&	0.000217	&	0.000197	&	0.000199	&	0.000235	&	0.000372		\\ \hline
Y Cyg	&	D	&	0.002855	&	0.002584	&	0.002618	&	0.003093	&	0.062096		\\ \hline
V380 Cyg	&	D	&	0.000770	&	0.000697	&	0.000706	&	0.000834	&	0.008785		\\ \hline
V453 Cyg	&	D	&	0.001865	&	0.001687	&	0.001709	&	0.002020	&	0.054265		\\ \hline
V477 Cyg	&	D	&	0.000731	&	0.000661	&	0.000670	&	0.000791	&	0.006614		\\ \hline
V1765 Cyg	&	D	&	0.001153	&	0.001044	&	0.001057	&	0.001250	&	0.006834		\\  \hline
57 Cyg	&	D	&	0.001321	&	0.001195	&	0.001211	&	0.001431	&	0.013870		\\ \hline
RU Mon	&	D	&	0.000993	&	0.000898	&	0.000910	&	0.001076	&	0.010160		\\ \hline
BM Mon	&	C	&	0.002941	&	0.002660	&	0.002695	&	0.003186	&	0.007309		\\ \hline
GM Nor	&	D	&	0.000902	&	0.000816	&	0.000827	&	0.000977	&	0.020653		\\ \hline
U Oph	&	D	&	0.001763	&	0.001595	&	0.001616	&	0.001910	&	0.078036		\\ \hline
V451 Oph	&	D	&	0.000961	&	0.000869	&	0.000881	&	0.001041	&	0.012036		\\ \hline
$\delta$ Ori	&	D	&	0.001729	&	0.001564	&	0.001585	&	0.001873	&	0.024907		\\ \hline
$\iota$ Ori	&	D	&	0.002067	&	0.001870	&	0.001895	&	0.002239	&	0.011973		\\ \hline
\hline
 \\
\end{tabular}
\label{omegap}
\end{table*}

\begin{table*}
\center
\addtocounter{table}{-1}
\caption{(continued)}
\begin{tabular}{|l|l|l|l|l|l|l|}
\hline
\hline
Name	&	Type	&	$\dot\omega_{GR}$	&	$\dot\omega_{BD}$	&	$\dot\omega_{ND}$	&	$\dot\omega_{f(R)}$	&	$\dot\omega_{Obs}$	\\ \hline \hline
FT Ori	&	D	&	0.000863	&	0.000781	&	0.000791	&	0.000935	&	0.006460		\\ \hline
AG Per	&	D	&	0.001614	&	0.001460	&	0.001479	&	0.001748	&	0.026467		\\ \hline
IQ Per	&	D	&	0.001142	&	0.001033	&	0.001046	&	0.001237	&	0.014451		\\ \hline
$\zeta$ Phe	&	D	&	0.001346	&	0.001218	&	0.001234	&	0.001458	&	0.037260		\\ \hline
KX Pup	&	D	&	0.000887	&	0.000802	&	0.000813	&	0.000961	&	0.012455		\\ \hline
NO Pup	&	D	&	0.001273	&	0.001151	&	0.001167	&	0.001379	&	0.033327		\\ \hline
VV Pyx	&	D	&	0.000516	&	0.000467	&	0.000473	&	0.000559	&	0.001417		\\ \hline
YY Sgr	&	D	&	0.000997	&	0.000902	&	0.000914	&	0.001080	&	0.008729		\\ \hline
V523 Sgr	&	D	&	0.000644	&	0.000583	&	0.000591	&	0.000698	&	0.011291		\\ \hline
V526 Sgr	&	D	&	0.000973	&	0.000880	&	0.000892	&	0.001054	&	0.012135		\\ \hline
V1647 Sgr	&	D	&	0.000769	&	0.000696	&	0.000705	&	0.000834	&	0.005465		\\ \hline
V760 Sco	&	D	&	0.001709	&	0.001546	&	0.001566	&	0.001851	&	0.042681		\\ \hline
AO Vel	&	D	&	0.001613	&	0.001460	&	0.001479	&	0.001748	&	0.027516		\\ \hline
EO Vel	&	D	&	0.000644	&	0.000582	&	0.000590	&	0.000697	&	0.003285		\\ \hline
HR 8384	&	D	&	0.001172	&	0.001060	&	0.001074	&	0.001270	&	0.031373		\\ \hline
HR 8800	&	D	&	0.001562	&	0.001413	&	0.001431	&	0.001692	&	0.023023	
	\\ \hline
\hline
 \\
\end{tabular}
\end{table*}

\begin{table*}
\center
\caption{Ratio of $\frac{\dot\omega_{Rel}}{\dot\omega_{Obs}}$ and $\frac{\dot\omega_{Rel}}{\dot\omega_{cl}}$  for different relativistic theories (GR,BD,ND,f(R)) }
\begin{tabular}{|l|l|l|l|l|l|l|l|l|l|}
\hline
\hline
Name	&	Type	&	$\frac{\dot\omega_{GR}}{\dot\omega_{Obs}}	$ &$	\frac{\dot\omega_{GR}}{\dot\omega_{cl}}$	&	$\frac{\dot\omega_{BD}}{\dot\omega_{Obs}}$	&	$\frac{\dot\omega_{BD}}{\dot\omega_{cl}}$	&	$\frac{\dot\omega_{ND}}{\dot\omega_{Obs}}$	&	$\frac{\dot\omega_{ND}}{\dot\omega_{cl}}$	&	$\frac{\dot\omega_{f(R)}}{\dot\omega_{Obs}}	$ &	$\frac{\dot\omega_{f(R)}}{\dot\omega_{cl}}$	\\ \hline  \hline 
BW Aqr	&	D	&	0.55	&	0.36	&	0.48	&	0.32	&	0.48	&	0.33	&	0.63	&	0.39	\\ \hline
V889 Aql	&	D	&	2.92	&	0.74	&	2.07	&	0.67	&	2.15	&	0.68	&	4.17	&	0.81	\\ \hline
V539 Ara	&	D	&	0.07	&	0.06	&	0.06	&	0.06	&	0.06	&	0.06	&	0.07	&	0.07	\\ \hline
HP Aur	&	D	&	0.27	&	0.21	&	0.24	&	0.19	&	0.24	&	0.20	&	0.30	&	0.23	\\ \hline
AS Cam	&	D	&	1.30	&	0.56	&	1.04	&	0.51	&	1.07	&	0.52	&	1.58	&	0.61	\\ \hline
EM Car	&	C	&	0.04	&	0.04	&	0.04	&	0.03	&	0.04	&	0.03	&	0.04	&	0.04	\\ \hline
GL Car	&	D	&	0.03	&	0.03	&	0.03	&	0.03	&	0.03	&	0.03	&	0.03	&	0.03	\\ \hline
HH Car	&	SD	&	1.10	&	0.52	&	0.90	&	0.47	&	0.92	&	0.48	&	1.31	&	0.57	\\ \hline
QX Car	&	D	&	0.14	&	0.12	&	0.12	&	0.11	&	0.12	&	0.11	&	0.15	&	0.13	\\ \hline
AR Cas	&	D	&	0.12	&	0.11	&	0.11	&	0.10	&	0.11	&	0.10	&	0.14	&	0.12	\\ \hline
OX Cas	&	D	&	0.04	&	0.04	&	0.03	&	0.03	&	0.04	&	0.03	&	0.04	&	0.04	\\ \hline
PV Cas	&	D	&	0.07	&	0.06	&	0.06	&	0.06	&	0.06	&	0.06	&	0.07	&	0.07	\\ \hline
KT Cen	&	D	&	0.07	&	0.07	&	0.06	&	0.06	&	0.07	&	0.06	&	0.08	&	0.07	\\ \hline
V346 Cen	&	D	&	0.07	&	0.07	&	0.06	&	0.06	&	0.06	&	0.06	&	0.08	&	0.07	\\ \hline
CW Cep	&	D	&	0.04	&	0.04	&	0.04	&	0.03	&	0.04	&	0.03	&	0.04	&	0.04	\\ \hline
EK Cep	&	D	&	0.70	&	0.41	&	0.60	&	0.37	&	0.61	&	0.38	&	0.81	&	0.45	\\ \hline
NY Cep	&	D	&	0.08	&	0.08	&	0.08	&	0.07	&	0.08	&	0.07	&	0.09	&	0.08	\\ \hline
$\alpha$ Crb	&	D	&	1.40	&	0.58	&	1.12	&	0.53	&	1.15	&	0.54	&	1.72	&	0.63	\\ \hline
Y Cyg	&	D	&	0.05	&	0.05	&	0.04	&	0.04	&	0.04	&	0.04	&	0.05	&	0.05	\\ \hline
V380 Cyg	&	D	&	0.10	&	0.09	&	0.09	&	0.08	&	0.09	&	0.08	&	0.10	&	0.09	\\ \hline
V453 Cyg	&	D	&	0.04	&	0.03	&	0.03	&	0.03	&	0.03	&	0.03	&	0.04	&	0.04	\\ \hline
V477 Cyg	&	D	&	0.12	&	0.11	&	0.11	&	0.10	&	0.11	&	0.10	&	0.14	&	0.12	\\ \hline
V1765 Cyg	&	D	&	0.20	&	0.17	&	0.18	&	0.15	&	0.18	&	0.15	&	0.22	&	0.18	\\ \hline
57 Cyg	&	D	&	0.11	&	0.10	&	0.09	&	0.09	&	0.10	&	0.09	&	0.12	&	0.10	\\ \hline
RU Mon	&	D	&	0.11	&	0.10	&	0.10	&	0.09	&	0.10	&	0.09	&	0.12	&	0.11	\\ \hline
BM Mon	&	C	&	0.67	&	0.40	&	0.57	&	0.36	&	0.58	&	0.37	&	0.77	&	0.44	\\ \hline
GM Nor	&	D	&	0.05	&	0.04	&	0.04	&	0.04	&	0.04	&	0.04	&	0.05	&	0.05	\\ \hline
U Oph	&	D	&	0.02	&	0.02	&	0.02	&	0.02	&	0.02	&	0.02	&	0.03	&	0.02	\\ \hline
V451 Oph	&	D	&	0.09	&	0.08	&	0.08	&	0.07	&	0.08	&	0.07	&	0.09	&	0.09	\\ \hline
$\delta$ Ori	&	D	&	0.07	&	0.07	&	0.07	&	0.06	&	0.07	&	0.06	&	0.08	&	0.08	\\ \hline
$\iota$ Ori	&	D	&	0.21	&	0.17	&	0.19	&	0.16	&	0.19	&	0.16	&	0.23	&	0.19	\\ \hline 
 \hline
 \\
\end{tabular}
\label{omegap-ratio}
\end{table*}

\begin{table*}
\center
\addtocounter{table}{-1}
\caption{continued}
\begin{tabular}{|l|l|l|l|l|l|l|l|l|l|}
\hline
\hline
Name	&	Type	&	$\frac{\dot\omega_{GR}}{\dot\omega_{Obs}}	$ &$	\frac{\dot\omega_{GR}}{\dot\omega_{cl}}$	&	$\frac{\dot\omega_{BD}}{\dot\omega_{Obs}}$	&	$\frac{\dot\omega_{BD}}{\dot\omega_{cl}}$	&	$\frac{\dot\omega_{ND}}{\dot\omega_{Obs}}$	&	$\frac{\dot\omega_{ND}}{\dot\omega_{cl}}$	&	$\frac{\dot\omega_{f(R)}}{\dot\omega_{Obs}}	$ &	$\frac{\dot\omega_{f(R)}}{\dot\omega_{cl}}$	\\ \hline  \hline 
FT Ori	&	D	&	0.15	&	0.13	&	0.14	&	0.12	&	0.14	&	0.12	&	0.17	&	0.14	\\ \hline
AG Per	&	D	&	0.06	&	0.06	&	0.06	&	0.06	&	0.06	&	0.06	&	0.07	&	0.07	\\ \hline
IQ Per	&	D	&	0.09	&	0.08	&	0.08	&	0.07	&	0.08	&	0.07	&	0.09	&	0.09	\\ \hline
$\zeta$ Phe 	&	D	&	0.04	&	0.04	&	0.03	&	0.03	&	0.03	&	0.03	&	0.04	&	0.04	\\ \hline
KX Pup	&	D	&	0.08	&	0.07	&	0.07	&	0.06	&	0.07	&	0.07	&	0.08	&	0.08	\\ \hline
NO Pup	&	D	&	0.04	&	0.04	&	0.04	&	0.03	&	0.04	&	0.04	&	0.04	&	0.04	\\ \hline
VV Pyx	&	D	&	0.57	&	0.36	&	0.49	&	0.33	&	0.50	&	0.33	&	0.65	&	0.39	\\ \hline
YY Sgr	&	D	&	0.13	&	0.11	&	0.12	&	0.10	&	0.12	&	0.10	&	0.14	&	0.12	\\ \hline
V523 Sgr	&	D	&	0.06	&	0.06	&	0.05	&	0.05	&	0.06	&	0.05	&	0.07	&	0.06	\\ \hline
V526 Sgr	&	D	&	0.09	&	0.08	&	0.08	&	0.07	&	0.08	&	0.07	&	0.10	&	0.09	\\ \hline
V1647 Sgr	&	D	&	0.16	&	0.14	&	0.15	&	0.13	&	0.15	&	0.13	&	0.18	&	0.15	\\ \hline
V760 Sco	&	D	&	0.04	&	0.04	&	0.04	&	0.04	&	0.04	&	0.04	&	0.05	&	0.04	\\ \hline
AO Vel	&	D	&	0.06	&	0.06	&	0.06	&	0.05	&	0.06	&	0.05	&	0.07	&	0.06	\\ \hline
EO Vel	&	D	&	0.24	&	0.20	&	0.22	&	0.18	&	0.22	&	0.18	&	0.27	&	0.21	\\ \hline
HR 8384	&	D	&	0.04	&	0.04	&	0.03	&	0.03	&	0.04	&	0.03	&	0.04	&	0.04	\\ \hline
HR 8800	&	D	&	0.07	&	0.07	&	0.07	&	0.06	&	0.07	&	0.06	&	0.08	&	0.07	\\ 
\hline
 \hline
 \\
\end{tabular}
\end{table*}
\begin{table*}
\center
\caption{
Internal second order structure  constants (ISC) $\log k_{2i}$  and $\log\bar{k}_{2Obs_{Th}}$ toghether $\log\bar{k_{2}}$ for different relativistic terms (GR,BD,ND,f(R))}
\begin{tabular}{|l|l|l|l|l|l|l|l|l|l|}

\hline
\hline

Name	&	Type	&	$\log k_{21}$	&	$\log k_{22}$	&	$\log\bar{k}_{2} $ &	$\log\bar{k}_{2_{ObsGR}}$	&	$\log\bar{k}_{2_{ObsBD}}$	&	$\log\bar{k}_{2_{ObsND}}$	&	$\log\bar{k}_{2_{Obsf(R)}}$\\ \hline \hline	
BW Aqr	&	D	&	-2.34	&	-2.354	&	-2.345	&	-2.245	&	-2.222	&	-2.225	&	-2.265	\\ \hline
V889 Aql	&	D	&	-2.392	&	-2.384	&	-2.389	&	-2.227	&	-2.121	&	-2.133	&	-2.348	\\ \hline
V539 Ara	&	D	&	-2.269	&	-2.209	&	-2.253	&	-2.290	&	-2.287	&	-2.287	&	-2.292	\\ \hline
HP Aur	&	D	&	-1.93	&	-1.66	&	-1.705	&	-2.638	&	-2.627	&	-2.629	&	-2.648	\\ \hline
AS Cam	&	D	&	-2.347	&	-2.339	&	-2.345	&	-3.027	&	-2.977	&	-2.983	&	-3.077	\\ \hline
EM Car	&	C	&	-2.172	&	-2.128	&	-2.154	&	-2.304	&	-2.302	&	-2.302	&	-2.305	\\ \hline
GL Car	&	D	&	-1.907	&	-1.909	&	-1.908	&	-1.880	&	-1.878	&	-1.879	&	-1.881	\\ \hline
HH Car	&	SD	&	-2.062	&	-2.055	&	-2.055	&	-4.458	&	-4.415	&	-4.420	&	-4.499	\\ \hline
QX Car	&	D	&	-2.331	&	-2.321	&	-2.326	&	-2.075	&	-2.070	&	-2.070	&	-2.080	\\ \hline
AR Cas	&	D	&	-2.221	&	-2.182	&	-2.218	&	-1.855	&	-1.850	&	-1.851	&	-1.860	\\ \hline
OX Cas	&	D	&	-2.394	&	-2.398	&	-2.396	&	-2.326	&	-2.324	&	-2.325	&	-2.327	\\ \hline
PV Cas	&	D	&	-2.343	&	-2.313	&	-2.327	&	-2.475	&	-2.473	&	-2.473	&	-2.478	\\ \hline
KT Cen	&	D	&	-2.339	&	-2.111	&	-2.225	&	-2.210	&	-2.207	&	-2.207	&	-2.213	\\ \hline
V346 Cen	&	D	&	-2.047	&	-2.034	&	-2.046	&	-2.356	&	-2.353	&	-2.354	&	-2.359	\\ \hline
CW Cep	&	D	&	-2.222	&	-2.305	&	-2.254	&	-2.102	&	-2.101	&	-2.101	&	-2.104	\\ \hline
EK Cep	&	D	&	-2.176	&	-2.095	&	-2.130	&	-2.045	&	-2.017	&	-2.020	&	-2.071	\\ \hline
NY Cep	&	D	&	-2.524	&	-1.497	&	-1.863	&	-1.486	&	-1.482	&	-1.483	&	-1.489	\\ \hline
$\alpha$ Crb	&	D	&	-1.968	&	-1.966	&	-1.968	&	-2.001	&	-1.947	&	-1.953	&	-2.055	\\ \hline
Y Cyg	&	D	&	-2.768	&	-2.116	&	-2.369	&	-1.970	&	-1.968	&	-1.969	&	-1.972	\\ \hline
V380 Cyg	&	D	&	-2.382	&	-2.133	&	-2.381	&	-3.046	&	-3.042	&	-3.042	&	-3.049	\\ \hline
V453 Cyg	&	D	&	-2.37	&	-2.214	&	-2.346	&	-2.264	&	-2.262	&	-2.263	&	-2.265	\\ \hline
V477 Cyg	&	D	&	-2.146	&	-2.137	&	-2.143	&	-2.247	&	-2.242	&	-2.243	&	-2.252	\\ \hline
V1765 Cyg	&	D	&	-2.227	&	-2.212	&	-2.227	&	-3.200	&	-3.192	&	-3.193	&	-3.207	\\ \hline
57 Cyg	&	D	&	-2.129	&	-2.149	&	-2.136	&	-2.305	&	-2.300	&	-2.301	&	-2.309	\\ \hline
RU Mon	&	D	&	-2.123	&	-2.218	&	-2.158	&	-2.213	&	-2.209	&	-2.209	&	-2.217	\\ \hline
BM Mon	&	C	&	-2.548	&	-2.47	&	-2.524	&	-4.459	&	-4.432	&	-4.435	&	-4.484	\\ \hline
GM Nor	&	D	&	-2.297	&	-2.255	&	-2.290	&	-2.619	&	-2.617	&	-2.617	&	-2.620	\\ \hline
U Oph	&	D	&	-2.423	&	-2.39	&	-2.410	&	-2.227	&	-2.226	&	-2.226	&	-2.228	\\ \hline
V451 Oph	&	D	&	-2.811	&	-2.015	&	-2.431	&	-2.481	&	-2.477	&	-2.478	&	-2.484	\\ \hline
$\delta$ Ori	&	D	&	-2.647	&	-2.065	&	-2.402	&	-3.409	&	-3.406	&	-3.407	&	-3.412	\\ \hline
$\iota$ Ori	&	D	&	-2.415	&	-2.396	&	-2.411	&	-3.076	&	-3.068	&	-3.069	&	-3.084	\\ 
\hline
\hline
 \\
\end{tabular}
\label{k2table}
\end{table*}

\begin{table*}
\center
\addtocounter{table}{-1}
\caption{continued}

\begin{tabular}{|l|l|l|l|l|l|l|l|l|l|}
\hline
\hline
Name	&	Type	&	$\log k_{21}$	&	$\log k_{22}$	&	$\log\bar{k}_{2}$ &	$\log\bar{k}_{2_{ObsGR}}$	&	$\log\bar{k}_{2_{ObsBD}}$	&	$\log\bar{k}_{2_{ObsND}}$	&	$\log\bar{k}_{2_{Obsf(R)}}$ \\ \hline \hline
FT Ori	&	D	&	-2.095	&	-2.113	&	-2.103	&	-2.335	&	-2.329	&	-2.330	&	-2.341	\\ \hline
AG Per	&	D	&	-2.271	&	-2.371	&	-2.305	&	-2.139	&	-2.137	&	-2.137	&	-2.142	\\ \hline
IQ Per	&	D	&	-2.264	&	-2.302	&	-2.273	&	-2.360	&	-2.356	&	-2.357	&	-2.363	\\ \hline
$\zeta$ Phe	&	D	&	-2.386	&	-2.389	&	-2.387	&	-2.222	&	-2.220	&	-2.220	&	-2.223	\\ \hline
KX Pup	&	D	&	-2.264	&	-2.313	&	-2.274	&	-2.360	&	-2.357	&	-2.357	&	-2.363	\\ \hline
NO Pup	&	D	&	-2.435	&	-2.435	&	-2.435	&	-2.309	&	-2.307	&	-2.308	&	-2.310	\\ \hline
VV Pyx	&	D	&	-2.413	&	-2.385	&	-2.399	&	-2.489	&	-2.466	&	-2.469	&	-2.511	\\ \hline
YY Sgr	&	D	&	-2.403	&	-2.391	&	-2.398	&	-2.285	&	-2.280	&	-2.281	&	-2.290	\\ \hline
V523 Sgr	&	D	&	-2.462	&	-2.419	&	-2.456	&	-2.725	&	-2.723	&	-2.723	&	-2.728	\\ \hline
V526 Sgr	&	D	&	-2.347	&	-2.361	&	-2.352	&	-2.360	&	-2.357	&	-2.357	&	-2.364	\\ \hline
V1647 Sgr	&	D	&	-2.446	&	-2.414	&	-2.432	&	-2.366	&	-2.359	&	-2.360	&	-2.372	\\ \hline
V760 Sco	&	D	&	-2.395	&	-2.395	&	-2.395	&	-2.200	&	-2.198	&	-2.198	&	-2.201	\\ \hline
AO Vel	&	D	&	-2.482	&	-2.482	&	-2.482	&	-2.249	&	-2.246	&	-2.247	&	-2.251	\\ \hline
EO Vel	&	D	&	-2.39	&	-2.115	&	-2.246	&	-2.539	&	-2.529	&	-2.530	&	-2.548	\\ \hline
HR 8384	&	D	&	-2.344	&	-1.746	&	-1.979	&	-2.097	&	-2.096	&	-2.096	&	-2.099	\\ \hline
HR 8800	&	D	&	-2.146	&	-2.208	&	-2.162	&	-2.591	&	-2.588	&	-2.589	&	-2.594	\\ \hline
 \hline
 \\
\end{tabular}
\end{table*}

\label{lastpage}

\end{document}